# Collective excitations in liquid D$_2$ confined within the mesoscopic pores of a MCM-41 molecular sieve

**Claudia Mondelli, Miguel A. González,**
*Institut Laue Langevin, B.P. 156x, 38042 Grenoble Cedex 9, France*

**Francesco Albergamo,**
*European Synchrotron Radiation Facility, B.P. 220, F-38043 Grenoble Cedex, France*

**Carlos Carbajo,**
*Departamento de Química Inorgánica, Facultad de Ciencias Químicas, Universidad Complutense, E-28049 Madrid, Spain*

**M. Jose Torralvo, Eduardo Enciso,**
*Departamento de Química Física I, Facultad de Ciencias Químicas, Universidad Complutense, E-28049 Madrid, Spain*

**F. Javier Bermejo,**
*C.S.I.C.- Department of Electricity and Electronics, University of the Basque Country, P.O. Box. 644, Bilbao 48080, Spain*

**Ricardo Fernández-Perea, Carlos Cabrillo,**
*Instituto de Estructura de la Materia, Consejo Superior de Investigaciones Cientificas, Serrano 123, Madrid 28006, Spain*

**Vincent Leon, Marie-Louise Saboungi**
*Centre de Recherche sur la Matiére Divisée, 1B rue de la Férollerie 45071, Orléans Cedex 2, France*

## ABSTRACT

We present a comparative study of the excitations in bulk and liquid D$_2$ confined within the pores of MCM-41. The material (Mobile Crystalline Material-41) is a silicate obtained by means of a template that yields a partially crystalline structure composed by arrays of nonintersecting hexagonal channels of controlled width having walls made of amorphous SiO$_2$. Its porosity was characterized by means of adsorption isotherms and found to be composed by a regular array of pores having a narrow distribution of sizes with a most probable value of 2.45 nm. The assessment of the precise location of the sample within the pores is carried out by means of pressure isotherms. The study was conducted at two pressures which correspond to pore fillings above the capillary condensation regime. Within the range of wave vectors where collective excitations can be followed up ($0.3 \leq Q \leq 3.0$ Å$^{-1}$), we found confinement brings forward a large shortening of the excitation lifetimes that shifts the characteristic frequencies to higher energies. In addition, the coherent quasielastic scattering shows signatures of reduced diffusivity.

# I.   INTRODUCTION

Within the last two decades, a great deal of interest has been focused onto the expectancy of new phenomena arising as a consequence of the reduction of the spatial scale of a liquid if confined within materials having pore sizes within the few nanometers range [1]. Particular interest merits the possibility of obtaining a steady liquid phase in a density and temperature range well beyond the stability limit of the bulk phase [2], [3]. Within this context, supercooling of liquid hydrogen has been a topic of active research for some decades now since it is expected that it will enhance its quantum effects and at some point it may enter the superfluid regime as conjectured some time ago [4]. A superfluid phase has been recently predicted as well for very-high-density regimes [5]. In the quest for supercooling liquid hydrogen down to temperatures as low as 4 K where the superfluid transition is expected to take place, confinement within porous media reported on a strong reduction of the freezing point from about 14 K for the bulk liquid down to about 8 K when confined in Vycor glass that is a tenuous disordered structure [6]. Further confinement within materials having pore sizes of the order of a nanometer such as zeolites was an obvious next step. However, strong interactions between the liquid and the pore walls seem to dominate the response [7], which preempts the formation of a superfluid. On the one hand, more recent results on new ordered materials, such as MCM-48 having pores of the order of two nanometers arranged in a regular three-dimensional array [8], showed that confinement within the material had no effect in depressing the freezing point below the triple point of the bulk liquid and therefore the results dashed hopes of progress along this route.

On the other hand, the now abundant data about the microscopics of liquids confined within porous materials have evidenced a rather complex phenomenology resulting from subtle details concerning the filling processes of liquids within the pores [9]. It thus seems obvious that a fundamental understanding of such processes is required to progress towards deep supercooling of the hydrogen liquids.

The case of liquid $^4$He has been studied in far more detail [2], [3], [10], [11] and the results have shown to be dependent upon fine details of the fluid-pore interactions. In fact, the whole phenomenology is shown to be dependent not only upon the material used as a confining medium but also on the amount of liquid filling the porous structure.

While there have been a good number of reports on the behavior of hydrogen confined in porous materials [6], [7], [8], [9], [12], the main emphasis has been focused onto the structure or single particle properties, leaving the effects of confinement on the collective dynamics largely unexplored. The issue is, however, of interest since, as demonstrated by work carried out for the helium liquids, these properties may exhibit entirely new phenomena resulting from the imposed restrictions upon the spatial scale. Here we report on a first study conducted using liquid $D_2$ confined within the well-known mesoporous MCM-41 [13] which has cylindrical pores with most probable dimensions of 2.45±0.3 nm, aiming to explore the effects of confinement on the collective excitations supported by the confined liquid. As a first study we choose to use deuterium rather than hydrogen to avoid the need of performing *ortho→para* catalysis cycles and also because of the rather similar spectra of collective excitations shown by both liquids [15].

## II. EXPERIMENT DETAILS

The MCM-41 sample was synthesized by mixing tetraethoxysilane (TEOS), ethanol (EtOH), and water (TEOS/EtOH/$H_2O$=1:4:1) for 1 h. The *p*H was adjusted with HCl to 1.2. This solution was added to another one formed by cetyltrimethylammonium bromide (CTAB), HCl, and $H_2O$ (CTAB/$H_2O$/HCl/TEOS=0.12:130: 9.2:1) and stirred for 2 h [16]. The resultant product was filtered, washed with water, and dried at 100 °C for 24 h. The white solid product was stirred with EtOH for 2 h, then the sample was filtered, washed with EtOH, and vacuum-dried at 100 °C. The sample was characterized by TEM (JEOL 2000FX electron microscope operating at 200 kV), low angle x-ray diffraction (Philips X'PERT-MPD diffractometer using Cu $K\alpha$ radiation), and nitrogen adsorption at 77 K (ASAP 2020 analyzer from Micromeritics). The TEM micrographs show ordered stripes with periodic length of about 2.45 nm (see Fig.1). The microstructure of the sample appears as formed by microdomains in different orientations and in some areas 2D-hexagonal order can be also observed. The images are consistent with a tubelike mesostructure characteristic of MCM-41 materials. The x-ray diffraction pattern shows a peak centered at 4.06 (2$\theta$) that can be assigned to (100) reflection of a hexagonal lattice with $a_0$=2.5 nm ($a_0 = 2d_{100}/\sqrt{3}$). The nitrogen adsorption isotherm (see Fig. 2) and the corresponding *t* plot indicate that mesopore filling occurs at relative pressure $P/P_0$=0.2. The BET area [17], calculated with the value 0.162 nm$^2$ for the molecular cross-sectional area of adsorbed nitrogen, was 1220 m$^2$ g$^{-1}$. The value of 20 m$^2$ g$^{-1}$ was obtained for the external area from the standard *t*-plot, with *t*, the statistical thickness of the adsorbed film, calculated by the Harkins and Jura equation. Pore size distribution, obtained by the BJH analysis from both adsorption and desorption branches of the isotherm, show a well-defined maximum for pore diameter of ~2.45 nm with a full width at half-maximum of ~0.3 nm. The Barret, Joyner, and Halenda (BJH) [18] cumulative pore was ~0.9 cm$^3$ g$^{-1}$.

The particular mesoscopic phenomena occurring in $D_2$ confined within the present material were investigated by direct measurement of the adsorption isotherm of 99.9% purity $D_2$ at *T*=19.6 K on 55.9 mg of the MCM-41 sample. A Baratron transducer with precision of 0.12% of the reading was used to determine the pressure while the volume of each steel vessel involved was known to a precision of better than 0.8%. To determine the $D_2$ amounts we used the equation of state proposed in [19].

The results shown in Fig. 2 depict an initial phase up to fillings $f \approx 18$ mmol g$^{-1}$ where $D_2$ is tightly bound to the media wall and forms one or two solid layer(s) as reported previously for liquid $^4$He on similar materials [11]. Further adsorption up to the small change in curvature of the diagram at $f \approx 27$ mmol g$^{-1}$ and $P/P_0 \approx 0.15$ is believed to form the so-called multilayer which is one liquid layer on top of the solid one(s). The small change in curvature is interpreted as the start of the capillary condensation which ends at $f \approx 40$ mmol g$^{-1}$, where the adsorption curve become essentially flat: small adsorption is then allowed up to pressures close to the saturated vapor pressure.

Inelastic neutron scattering measurements were carried out using the thermal time-of-flight spectrometer IN4 hosted by the Institut Laue Langevin (Grenoble, France). The instrument configuration was chosen in order to cover a usable range in momentum transfers between 0.3 and 3.5 Å$^{-1}$ as dictated by previous experience [14]. Thus data were collected using a wavelength of 1.3 Å, corresponding to an incident energy $E_0$=48.4 meV and resulting in a resolution in energy transfers $\Delta\omega \approx 2$ meV (FWHM) at the elastic peak, as measured using a vanadium standard. Small variations in resolution within all the explored $Q$ ranges were found not to exceed 2.5%.

Deuterium was condensed *in situ* into a cylindrical container (either empty or filled with MCM-41) of diameter 14.9 mm. The pressure was controlled continuously all through the condensation process by means of a Baratron transducer connected to our gas transfer line.

In addition to the standard detector bank covering scattering angles between $13° \leq 2\theta \leq 120°$, we collected data also using the small-angle multidetector which covers the range between $3° \leq 2\theta \leq 9°$. However, instrumental noise hindered the use of the region below 0.3 Å$^{-1}$, while, due to the gap between the small-angle multidetector and the first standard detector, no useful data could be collected either between $Q$=0.7 Å$^{-1}$ and $Q$=1.2 Å$^{-1}$.

## III. RESULTS

Measurements were performed on a bulk liquid sample of $D_2$ at $T$=20 K, on $D_2$ + MCM-41 at $T$=20 K and two partial pressures approximately corresponding to $P/P_0$=0.32 and $P/P_0$=0.94 as referred to above, and finally two background runs using the empty container and the container loaded with MCM-41 before the $D_2$ loading. Data correction procedures were implemented within the LAMP ILL data analysis package to convert the measured intensities into double differential cross sections. The two chosen partial pressures mark the beginning and the end of the almost flat section of the adsorption isotherm after capillary condensation has taken place. Therefore, both of them correspond to nearly complete filling (about 92% and 99%, respectively) but disparate pressure conditions. Analysis of both datasets showed relatively small differences and consequently we will here report on data measured at the lower partial pressure.

A sample of the fully corrected spectra [22] for the bulk, confined liquid as well as the confining medium is shown in Fig. 3 for three selected momentum transfers. Their values are chosen to correspond to regions marking the initial riseup of the dispersion, that about the maxima of the dispersion curve and that corresponding to wave vectors where the static $S(Q)$ shows its maximum. The details pertaining the modeling of the total intensities have been given previously [14]. To recall it briefly, the total intensity is modeled as

$$I(Q,\omega) = [a_1 S_{coll}(Q,\omega) + a_2 S_2(Q,\omega)] \otimes R(\omega) \qquad (1)$$

where $S_{coll}(Q,\omega)$ stands for the collective response that is modeled as a damped harmonic oscillator characterized by an excitation frequency $\Omega_Q = \sqrt{\omega_Q^2 + \Gamma_Q^2}$ defined in terms of a bare frequency $\omega_Q$ renormalized by the friction term $\Gamma_Q$ as well as an excitation strength. Notice that $\Omega_Q$ can be interpreted as the physical frequency for a propagating excitation if its associated friction term does not become large compared to the $\omega_Q$ bare frequency. However, as we shall see below, the bulk liquid for $Q$ values well beyond 1 Å$^{-1}$ [14] and the confined liquid within all the explored range show large friction terms, which in turn make $\Omega_Q$ a parameter needed to quantitatively compare the data. Under such circumstances the motions represented by such a quantity are better thought of as being heavily damped (or overdamped) density oscillations rather than vibrational motions having a well-defined frequency. At any rate such parameters constitute useful tools to quantify the effects of confinement and therefore are used as descriptors for such phenomena.

The symbols $a_1$ and $a_2$ in Eq. (1) are weight factors left as adjustable. The contribution $S_2(Q,\omega)$ here comprises quasielastic scattering of both coherent and incoherent nature together with terms to account for rotational contributions for molecular deuterium (barely visible at the scales drawn in Fig. 3). A separation of coherent and incoherent quasielastic components as done previously [14] was not attempted here due to relatively low resolution in energy transfers [specified by the resolution function $R(\omega)$.

Data depicted in Fig. 3 show that for comparable momentum transfers the response of confined $D_2$ is substantially broader than that for the bulk liquid. The contribution from MCM-41 that is taken off after background subtraction comprises a strong elastic peak together with a weak but non-negligible excitation spectrum.

A quantitative comparison of results corresponding to the parameters $\Omega_Q$ and the associated friction (damping) terms is provided by plots shown in Fig. 4. The data for the confined liquid shown in the upper frame of Fig. 4 display consistently larger values that those corresponding to the bulk liquid. The latter are found to be in good agreement with those previously reported [14] which correspond to measurements using several different instruments. A glance at the lower frame of the figure just referred to suggests that the increase in excitation frequencies, which in part may be attributed to an increase in the elastic constant resulting from confinement, mostly results from a far larger damping term. To put things into numbers and with the proviso in mind about the physical meaning of $\Omega_Q$ commented on above, we have parametrized the dispersion curves using [20]

$$\Omega_Q = c_0 Q \left( 1 + \gamma Q^2 \frac{1-(Q/Q_a)^2}{1+(Q/Q_b)^2} \right) \qquad (2)$$

where $c_0$ stands for the hydrodynamic value of the sound velocity, $Q_a$ and $Q_b$ are constants[21] and the parameter governs the kinematics of phonon decay processes. The estimates for $c_0$ and $\gamma$ came out to be of $1139\pm120$ m s$^{-1}$ and $4.33\pm0.54$ for the bulk liquid and $1291\pm193$ m s$^{-1}$ and $7.32\pm0.69$ for the confined fluid. The numerical results just quoted point towards a somewhat stiffer liquid when confined and also indicate a significant increase in the probability of excitation decay through large angle decays [20]. In turn, the damping terms below $\sim1$ Å$^{-1}$ can be well approximated by $\Gamma_Q = D_\Gamma Q^\alpha$ having the exponent values of $1.05\pm0.11$ and $1.27\pm0.19$ that are well below that of $\alpha=2$ expected on hydrodynamics grounds while the $D_\Gamma$ prefactor is found to increase from $6.9\pm0.5$ up to $12.7\pm1.4$ as a result of confinement.

The minimum of $\Omega_Q$ versus wave vector curves is well approximated by a parabola in terms of $\Omega_Q = \Delta + (Q - Q_R)^2/W$ where $\Delta$ stands for the gap that increases from $2.79\pm0.05$ meV up to $4.86\pm0.12$ meV as the liquid is confined whereas the position of the minima as given by $Q_R=2.02\pm0.005$ barely shows any measurable change upon confinement (i.e., within error bars), and the same applies to the width term

$W$=0.072±0.003. The behavior of $\Omega_Q$ about such wave vectors is reminiscent of the Landau parabola used to describe the wave-vector dependence of frequencies about the roton minimum in superfluid $^4$He. However, due to the large friction terms, the three parameters describing such a curve are here employed as a device to quantify the effects of confinement. The results show that $Q_R$, which is basically given by the position where the maximum of the static structure factor is located, is hardly affected by confinement and consequently from here we infer that the density of the confined liquid must not differ largely from that of the bulk sample. As regards the gap term $\Delta$, the significant variation upon confinement is here understood as arising from a reduction in height of the main peak of $S(Q)$ due to the reduction of the spatial scale.

Significant differences are also found in the quasielastic part of the spectrum. As referred to above, the quantity here considered comprises both incoherent and coherent scattering contributions. The former is expected to dominate the response at low wave vectors ($Q$≤1 Å$^{-1}$) whereas the latter should significantly contribute at larger momentum transfers.

Data for the bulk liquid shown at the lower wave vectors (five data points) are well accounted for in terms of a Fickian (incoherent) diffusion law using a value for the self-diffusion coefficient of 3.34x10$^{-5}$ cm$^2$ s$^{-1}$ which favorably compares with the estimate of 3.6x10$^{-5}$ cm$^2$ s$^{-1}$ derived from previous studies using higher resolution in energy transfers. Above 1 Å$^{-1}$ significant deviations from hydrodynamic diffusion are expected mostly due to the mounting importance of coherent effects. The latter can be accounted for on a semi-quantitative basis using the approach due to Cohen *et al.* [23] developed for hard-sphere liquids but known to constitute a good approximation to describe the coherent quasielastic linewidth.

$\Delta\omega_{coh}(Q)$ is the maxima of $S(Q)$ for some complex liquids [24]. The expression derived in Ref. 23 reads

$$\Delta\omega_{coh}(Q) = \frac{D_E Q^2}{S(Q)[1 - j_0(Q\sigma) + 2j_2(Q\sigma)]} \qquad (3)$$

where $D_E$ is the hard-sphere self-diffusion coefficient, $j_0$, $j_2$ are spherical Bessel functions, $\sigma$ characterizes the size of the fluid particles, and $S(Q)$ is the static structure factor that shows a sharp maximum at $Q\sigma \approx 2\pi$. The latter quantity has been measured under conditions reported in a previous study where the structure factors of liquid parahydrogen and ortho-deuterium were measured by means of low-energy neutron diffraction [25] and shows a maximum at $Q_p$=2.139 Å$^{-1}$ from where one infers a suitable value for $\sigma$=2.937 Å. As shown in Fig. 3, the approximation works remarkably well within 1.5 Å$^{-1}$≤$Q$≤3.0 Å$^{-1}$ if the value for $D_E$ is identified with that for the self-diffusion coefficient just referred to above. In other words, Eq. 3 aptly describes the wave-vector dependence of the quasielastic width within the alluded range of wave vectors for the bulk liquid without leaving any free parameter. In contrast, such an approximation fails to reproduce the data for the confined liquid, even at a qualitative level (Fig. 5).

As also shown in the figure, the effects of confinement are dramatic, largely reducing the diffusion coefficient. Because of the strong low-to-middle angle scattering due to the confining material [9], reliable quasielastic data at the lowest wave vectors could not be derived, thus hampering the estimation of the diffusion coefficient.

However, what merits attention is the fact that for the filling conditions here reported, the diffusion coefficients appear to be largely reduced compared to that for the bulk liquid. Also, the shape of $\Delta\omega_{q.el}(Q)$ comes to be significantly different from that for the bulk. The presence of a maximum rather than a minimum about $Q_p$ is suggestive of the operation of jumplike diffusion motions rather than long-range translational diffusion. A more detailed investigation of this particular feature clearly calls for a study using higher energy resolution.

## IV. CONCLUSION

In summary, the results reported on here depict the action of strong interactions between adsorbed layers on the pore surfaces and the remaining liquid. At the pore filling considered here in more detail, which corresponds to complete capillary condensation but the partial pressure ($P/P_0$=0.32) being well below the vapor pressure of the bulk liquid, we found such interactions to substantially reduce the lifetime of the collective excitations, which translates into a large increase in linewidth as well as a reduced diffusivity with respect to the bulk liquid. Increasing the pressure up to a value close to the vapor pressure of the bulk liquid, $P/P_0$=0.94, shows no clear difference with respect to the state studied here. The origin of such phenomena can be thought of as arising from friction effects of the flowing liquid against the adsorbed solidlike layers. Further studies on the dependence of the excitation spectra at lower pore fillings are now envisloned.

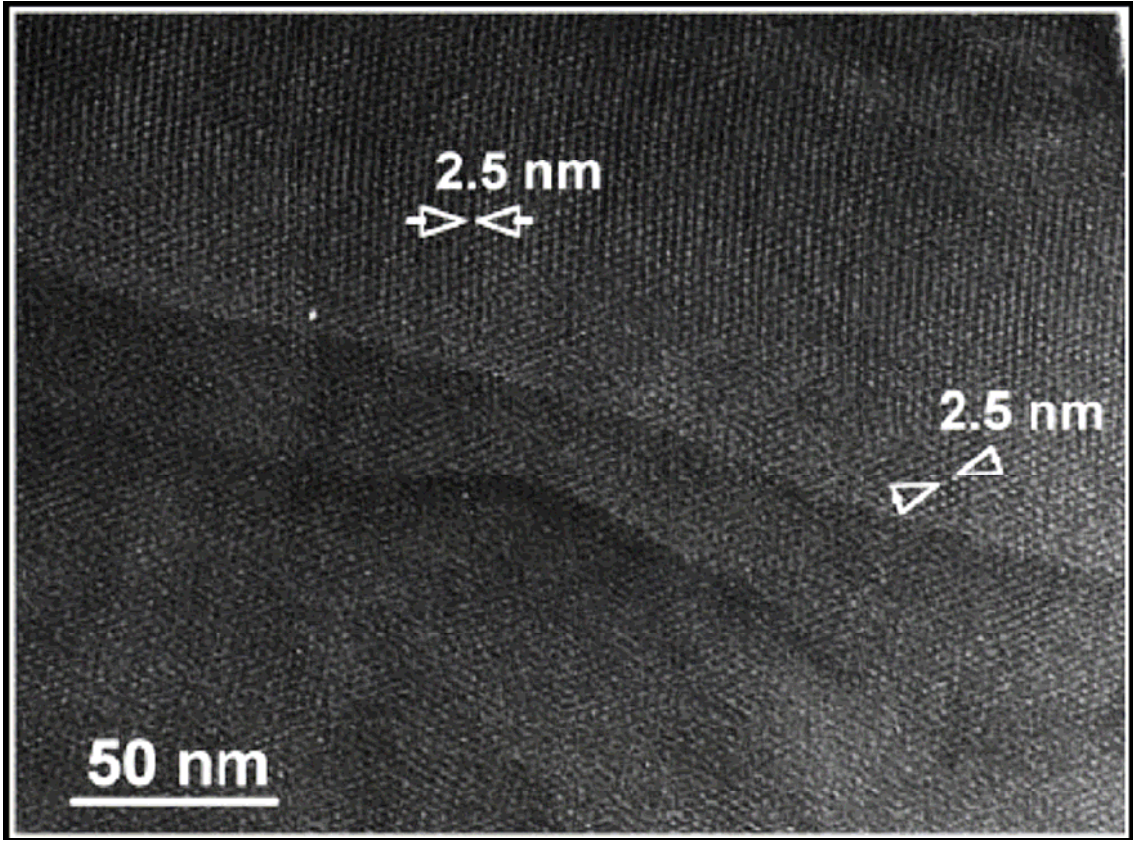

FIG. 1. TEM micrograph of the MCM-41 material prepared as explained in the text.

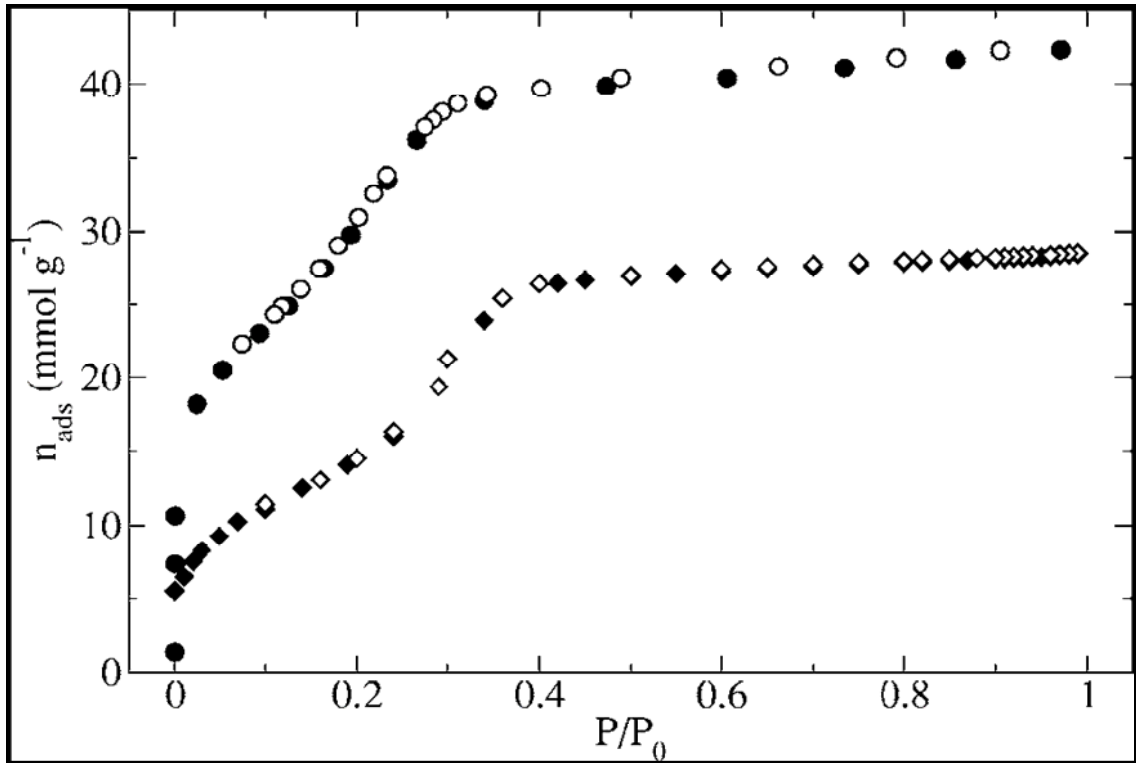

FIG. 2. Adsorption isotherm diagrams of $D_2$ at $T=19.6$ K (circles) and $N_2$ at $T=77$ K (diamonds). The abscissa corresponds to the ratio of the measured pressure to that of the saturated vapor pressure at such temperature. Filled and open symbols depict adsorption and desorption measurements, respectively. Error bars are smaller than the symbols size.

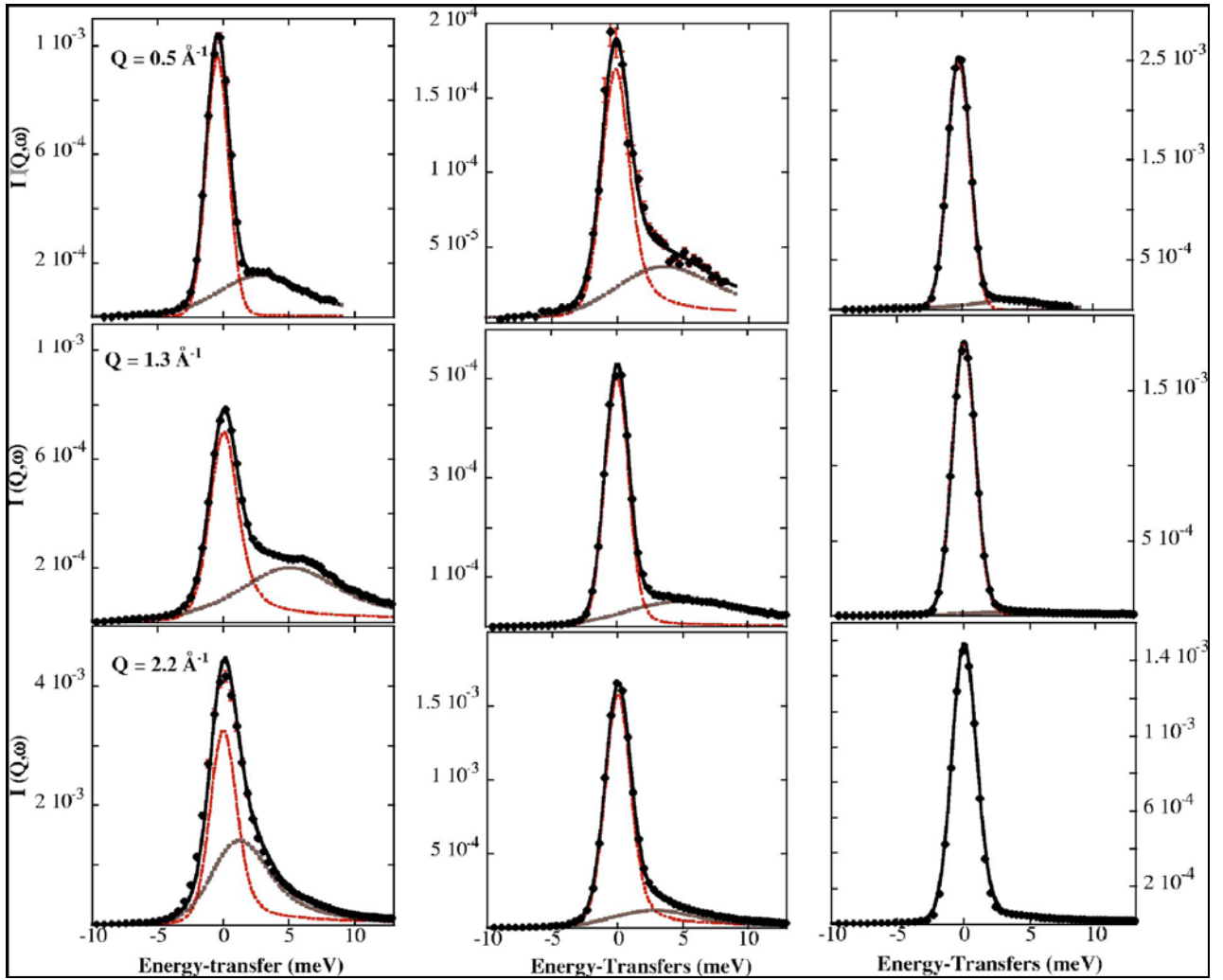

FIG. 3. (Color online) A sample of spectra corresponding to bulk liquid $D_2$ (left frames), liquid $D_2$ confined within MCM-41 pores and $P/P_0$=0.32 (middle frames), and those arising from the empty MCM-41 (right frames). Solid symbols depict experimental data, the thick solid line shows the model fit. The dotted line centered at zero freqeuencies shows the quasielastic component and the dashed line shows the spectrum from collective excitations.

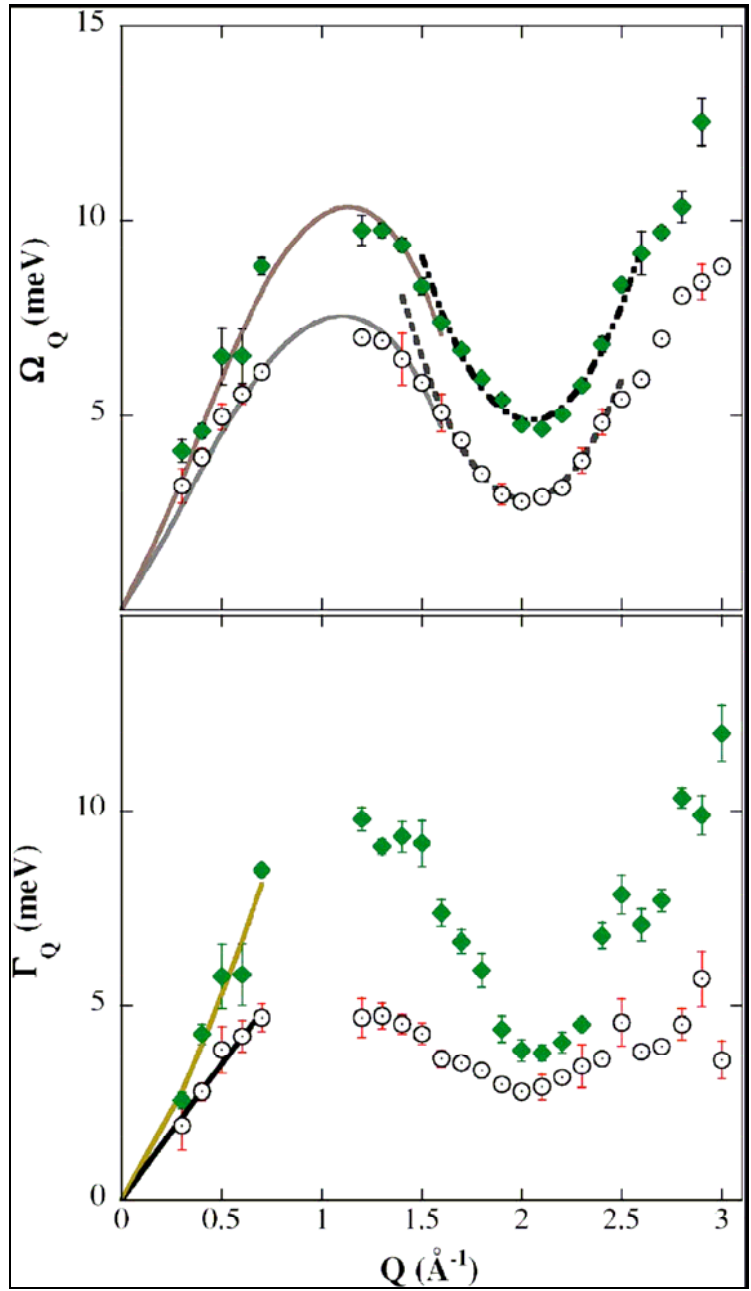

FIG. 4. (Color online) The upper frame depicts the measured $\Omega_Q$ for the bulk liquid (open circles) and the confined fluid solid lozenges. Solid lines correspond to parametric expressions for the dispersion (see text) and the dashed lines about the minima of both curves show fits to parabolae (see text). The lower frame shows the damping terms (same symbols as above). The solid line show fits to emphasize the non-hydrodynamic dependence with wave vector of the damping terms (see text).

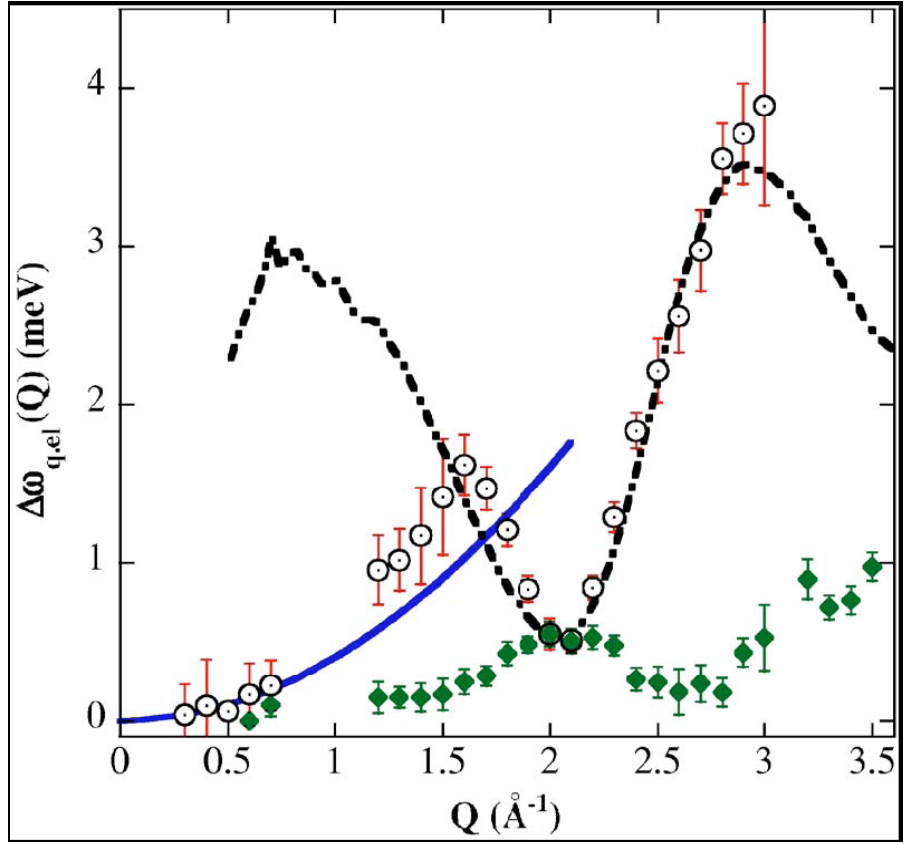

FIG. 5. (Color online)Wave-vector dependence of the quasielastic linewidths $\Delta\omega_{q.el}(Q)$ for the bulk (open circles) and confined (filled lozeges) liquids. The solid line represents Fickian diffusion following $\Delta\omega_{q.el}(Q)=D_sQ^2$ that yields a rough estimate for the selfdiffusion coefficient. The dashed line depicts an approximation to calculate the linewidth of the coherent part of the quasielastic spectrum [23] (see text).